# STEREOVISION IMAGE PROCESSING FOR PLANETARY NAVIGATION MAPS WITH SEMI-GLOBAL MATCHING AND SUPERPIXEL SEGMENTATION

Yan-Shan Lu[a*], Miguel Arana-Catania[b], Saurabh Upadhyay[c], Leonard Felicetti[d]

[a] *Faculty of Engineering and Applied Sciences, Cranfield University, United Kingdom, yan-shan.lu.609@cranfield.ac.uk*
[b] *Digital Scholarship at Oxford, University of Oxford, United Kingdom, humd0244@ox.ac.uk*
[c] *Faculty of Engineering and Applied Sciences, Cranfield University, United Kingdom, saurabh.upadhyay@cranfield.ac.uk*
[d] *Faculty of Engineering and Applied Sciences, Cranfield University, United Kingdom, leonard.felicetti@cranfield.ac.uk*
[*] *Corresponding author*

**ABSTRACT**

Mars exploration requires precise and reliable terrain models to ensure safe rover navigation across its unpredictable and often hazardous landscapes. Stereoscopic vision serves a critical role in the rover's perception, allowing scene reconstruction by generating precise depth maps through stereo matching. State-of-the-art Martian planetary exploration uses traditional local block-matching, aggregates cost over square windows, and refines disparities via smoothness constraints. However, this method often struggles with low-texture images, occlusion, and repetitive patterns because it considers only limited neighbouring pixels and lacks a wider understanding of scene context. This paper uses Semi-Global Matching (SGM) with superpixel-based refinement to mitigate the inherent block artefacts and recover lost details. The approach balances the efficiency and accuracy of SGM and adds context-aware segmentation to support more coherent depth inference. SGM provides an initial dense disparity map with subpixel accuracy by aggregating matching costs along multiple scanline paths, reducing the reliance on a fixed window size. However, while SGM handles smooth regions well, it can still introduce artefacts around object boundaries due to its pixel-level optimisation. To address this, superpixels are used to partition the image into perceptually uniform segments. Each superpixel is approximated as a planar surface, and disparities are refined to better conform to these planar constraints, improving alignment with object boundaries, reducing noise, and preserving fine geometric details. The proposed method has been evaluated in three datasets with successful results: In a Mars analogue, the terrain maps obtained show improved structural consistency, particularly in sloped or occlusion-prone regions. Large gaps behind rocks, which are common in raw disparity outputs, are reduced, and surface details like small rocks and edges are captured more accurately. Another two datasets, evaluated to test the method's general robustness and adaptability, show more precise disparity maps and more consistent terrain models, better suited for the demands of autonomous navigation on Mars, and competitive accuracy across both non-occluded and full-image error metrics. This paper outlines the entire terrain modelling process, from finding corresponding features to generating the final 2D navigation maps, offering a complete pipeline suitable for integration in future planetary exploration missions.

**Keywords:** Planetary Navigation, Semi-Global Matching, Superpixel Segmentation, Disparity Estimation, Terrain Mapping

## 1. INTRODUCTION

As Mars exploration missions continue to advance, accurate and reliable terrain models and navigation maps have become essential to ensure the rover can traverse safely in unpredictable landscapes [1]. Stereo vision plays a pivotal role in generating these models by reconstructing 3D information from rectified image pairs. The Mars exploration environment is often characterised by low texture, occlusion, and repetitive patterns, making it challenging to generate high-precision depth maps using stereo vision. This prompts us to explore new methods based on advanced algorithms to improve the accuracy of depth estimation and preserve terrain details. To the best of our knowledge, state-of-the-art Martian exploration techniques [2, 3] have predominantly adopted traditional local block-matching methods [4]. These methods employed local cost aggregation over square windows, although effective under many conditions, they often fail to capture reliable depth information in challenging planetary scenarios [5].

Recent advancements have turned to semi-global matching (SGM) [6] because of its ability to balance computational efficiency and accuracy. SGM simplifies the 2D global energy minimisation problem by performing 1D scanline optimisation along multiple directions. SGM's ability to produce accurate depth maps with manageable computational demands makes it a suitable choice for space exploration missions, where computing resources and robustness are critical. Recent works outlined the application of semi-global matching in generating dense disparity maps in the simulated lunar environment [7], built a coarse-to-fine pyramid framework with SGM to improve the accuracy of stereo matching for lunar rovers [8], and discussed methods used for rock measurement emphasis-ing SGM's effectiveness in handling the challenging conditions on the Martian surface during China's Tianwen-1 mission [9]. SGM has been widely used in real-time applications with proven qualities; however, it still faces issues in weak textures, sharp depth discontinuities, and loss of detail in occluded regions [10].

Moreover, alternative strategies that incorporate superpixel segmentation have emerged. Superpixels are perceptually meaningful regions in the image that follow object boundaries more closely than the rigid pixel structure, reducing redundancy and complexity of image processing tasks [11]. Many studies have shown the advantages of using superpixels as the basic processing unit [12–14]. More recently, [15] built a coarse-to-fine hierarchical framework based on a global matching method and superpixels to reduce runtime and complexity for lunar scenes. In addition, superpixel-based refinement methods perform plane fitting within each superpixel; these approaches improve alignment with object boundaries and enhance overall depth estimation. [16] over-segmented the image and labelled each region plane, and used global energy optimisation to solve the discontinuity problem in local matching. The state-of-the-art work [17] proposed a disparity estimation method based on optimised Census transform and superpixel refinement to improve the problem of weak texture and occlusion in lunar scenes.

In this paper, we propose a disparity refinement method that integrates SGM with superpixel-based segmentation to enhance overall disparity quality, providing a more reliable terrain model for rover navigation. The focus of the study is primarily on stereo matching, which is crucial for 3D reconstruction and navigation safety. Our work makes the following contributions:

- Proposing a novel disparity estimation method that integrates SGM with superpixel-based refinement to mitigate block artefacts and enhance overall quality.

- Implementing a robust plane-fitting approach within superpixels based on RANSAC [18] to align disparity estimates with true object boundaries.

- Validating the approach on planetary datasets analogous to Mars, demonstrating improved consistency and precision in the generated terrain models.

The remainder of the paper is organised as follows. Section 2 overviews the terrain modelling pipeline for planetary exploration. Section 3 details our proposed methodology in disparity estimation. Section 4 details further processes for 3D reconstruction. Section 5 presents experimental results and performance evaluations, and Section 6 concludes the paper.

## 2. TERRAIN MODELLING PIPELINE USING STEREO VISION

In this section, we give an overview of the stereo vision-based terrain modelling pipeline [2, 19]. Terrain modelling for rover navigation, as illustrated in Fig. 1, is a multi-step process that requires both image and point cloud processing to generate a navigation map. It follows these steps: (a) The process inputs are rectified stereo image pairs from the rover cameras; images whose corresponding points lie on the same image rows. (b) Pixel-wise disparities are then estimated by searching for corresponding features between the two images, resulting in a dense disparity map—a process known as stereo matching [5]. (c) Each pixel in the disparity map is subsequently re-projected into 3D space using the camera's intrinsic and extrinsic parameters, yielding a point cloud that represents the spatial distribution of terrain features. (d) This point cloud is further processed through voxelisation, which discretises the continuous set of 3D points into a structured grid of volumetric elements (voxels), making it more computationally efficient for terrain analysis. The voxelised representation is then used to generate a 2.5D digital elevation/terrain model (DEM/DTM) [1], where each grid cell represents an elevation measure and terrain characteristics. (e) Finally, the DEM/DTM is simplified into a 2D top-down binary navigation map that classifies regions as safe or unsafe based on terrain safety conditions, which is used to make local driving decisions [2, 19].

## 3. PROPOSED DISPARITY ESTIMATION METHOD

The proposed method is a disparity refinement approach that integrates Semi-Global Matching (SGM) with superpixel-based segmentation to improve depth estimation accuracy and boundary adherence. A flowchart is given in Fig. 2, which includes the following steps: (1) SGM is employed as the initial disparity estimation step to achieve subpixel accuracy and dense disparity maps. (2) we incorporate superpixel-based refinement using a plane-fitting approach based on RANSAC. (3) Post-processing and occlusion handling are applied to obtain the final disparity map. These three stages are described separately below.

### 3.1 Initial Disparity Map

The initial disparity map is computed using Semi-Global Matching (SGM) [6]. SGM approximates a global optimisation by aggregating matching costs along multiple independent 1D paths through the image rather than performing a full 2D global optimisation. The idea is to propagate the matching costs from neighbouring pixels while enforcing smoothness in the disparity map. Smoothness is encouraged by penalising changes in disparity be-tween neighbouring pixels. The goal is to compute the optimal disparity map $D$ by minimising the energy function [6]:

$$E(D) = \sum_{p} C(p, D_p) + \sum_{q \in N_p} P_1 \, \text{T}[|D_p - D_q| = 1] + \sum_{q \in N_p} P_2 \, \text{T}[|D_p - D_q| > 1] \quad [1]$$

where $C(p, D_p)$ is the data cost, representing how well each pixel $p$ matches disparity $D_p$, the second and third terms are the smoothness terms, which penalise large disparity differences for all pixels $q$ in the neighbourhood $N_p$ of $p$ to enforce local smoothness. $P_1$ penalises small disparity changes (i.e. 1 pixel), encouraging slight variations for slanted or curved surfaces, and $P_2$ is a larger penalty which penalises larger jumps to preserve discontinuities at object boundaries. The penalties can be tuned to control how strongly the algorithm enforces smoothness.

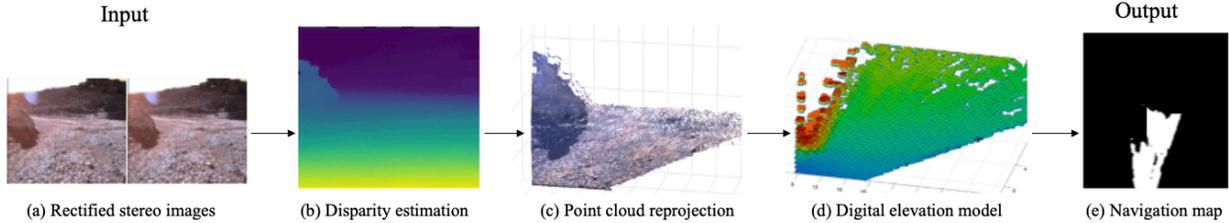

Fig. 1. Overview of the terrain modelling pipeline. Inputs of rectified stereo image pairs (a) are used to estimate a dense disparity map (b), reprojecting the disparities into 3D space to obtain a point cloud (c), followed by voxelisation to obtain a DEM (d), and a final top-down binary navigation map (e).

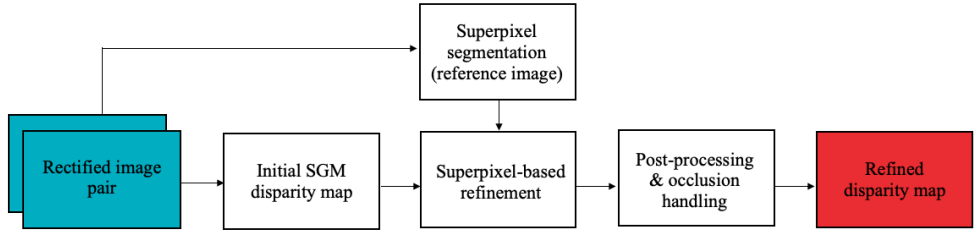

Fig. 2. Flowchart of our proposed method in disparity estimation. Coloured for the input (blue) and output (red).

SGM simplifies the problem by performing 1D scanline optimisation along multiple directions (typically 8). For each direction, the algorithm sums the matching costs from a sequence of neighbouring pixels, penalising large disparity changes along that direction. The final disparity map is determined by selecting the disparity with the minimum aggregated cost [6].

To optimise the initial estimates, left-right consistency checking and subpixel interpolation are conducted in the algorithm. Left-right consistency checking helps filter out unreliable estimates, especially in occluded areas, by comparing corresponding disparity maps from both stereo directions [5]. Parabolic interpolation is used to achieve subpixel precision, which refines the discrete disparity values into continuous ones for better depth resolution [6].

Fig. 3 (a) shows a sample output of the initial disparity map with a comparison of the original scene.

*3.2 Superpixel-Based Refinement*

To mitigate the artefact issues observed in Fig. 3(a), we incorporate superpixel-based refinement based on the initial disparity computation. Instead of refining disparity values at the pixel level, the image is segmented into superpixels using Simple Linear Iterative Clustering (SLIC) [20]. Each superpixel is treated as a piecewise planar surface, and its disparities are refined using a plane-fitting approach based on RANSAC [18].

**Superpixel Segmentation.** We segment the reference images into superpixels using the SLIC algorithm [20]. SLIC groups pixels in an image into compact superpixels using k-means clustering, in which the pixel dissimilarity metric depends on colour and image coordinate distances. Compactness of the superpixels can be adjusted as needed. SLIC stands out among superpixel segmentation methods for its simplicity and efficiency, particularly in terms of computational speed, segmentation quality, and adherence to image boundaries [20], making it a preferred choice in many real-world computer vision tasks requiring real-time processing or resource constraints [15], such as Mars rover navigation systems. An example of the resulting superpixels in one of the tested images is shown in Fig. 3 (b).

**Plane-Fitting in Segments.** To refine disparity estimates, we fit a 3D plane to the disparity values within each superpixel using the RANSAC (Random Sample Consensus) algorithm [18]. The core assumption behind this segmentation approach is that disparity discontinuities occur only at segment boundaries, allowing each superpixel to be modelled as a planar surface [21]. RANSAC is employed to estimate plane parameters robustly, making it resistant to noise and outliers. Following the method outlined in [21], plane-fitting enforces planar constraints within each segment, refining the initial disparity map while preserving structural integrity. To maintain accuracy, only non-occluded regions undergo refinement, preventing erroneous disparity adjustments in ambiguous areas. By leveraging RANSAC, the approach effectively mitigates outliers, preventing unreliable disparity estimates from distorting the plane-fitting process. This results in a smoother disparity map with reduced noise and artefacts, ensuring that depth discontinuities align with real scene structures rather than being overly smoothed.

*3.3 Post-Processing & Occlusion Handling*

To address noise and small fragmented errors caused by dust, lighting variations, or camera calibration issues, global refinement across neighbouring segments is crucial for maintaining consistency in disparity estimation [16]. Occluded regions, areas visible to only one camera, are inferred using Weighted Least Squares (WLS) filtering [11]. WLS is particularly effective for occlusion handling as it smooths unreliable regions while preserving object boundaries. Unlike simple interpolation, WLS assigns weights based on both spatial proximity and disparity confidence, ensuring that occluded areas are filled using nearby reliable data rather than arbitrary smoothing. Additionally, a bilateral filter is applied, leveraging both spatial closeness and photometric similarity to maintain fine details [11]. This combination prevents over-smoothing and helps retain depth discontinuities in the final disparity maps.

As shown in Fig. 3, the colour-coded refined disparity map exhibits significantly fewer artefacts, especially around object edges (previously marked as invalid in dark red). Overlaying Canny edges further illustrates that the proposed method improves disparity alignment with object boundaries, effectively handling occlusions while preserving structural details.

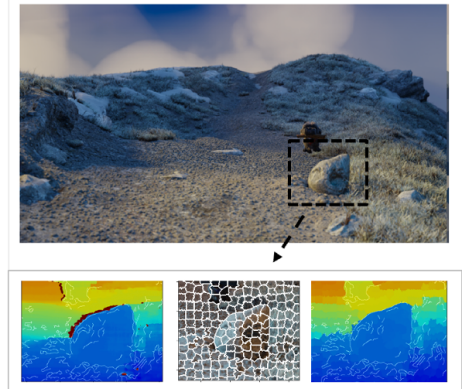

Fig. 3. Sample output from a scene in the Spring dataset [22]. Top row: original image. Bottom row is a zoom-in view, from left to right: (a) Initial disparity map, containing occlusions (marked as invalid in dark red), (b) superpixel segmentation, (c) refined disparity map using the proposed approach, adhering better to the object boundaries.

## 4. TERRAIN RECONSTRUCTION & NAVIGATION MAP

This section outlines the additional steps undertaken to build terrain models and create the final navigation maps based on the obtained disparity maps.

**Depth Estimation.** The disparity map obtained from stereo matching is used to reconstruct the 3D structure of the scene through triangulation, transforming 2D pixel coordinates $(u, v)$ into real-world 3D points $(X_c, Y_c, Z_c)$. The recovered depth (distance from the camera) of each point is computed based on the geometric relationship between the stereo cameras [5]. The output is a 3D point cloud, representing the spatial distribution of terrain features.

**Outlier Removal & ROI Limitation.** To clean noisy point clouds caused by disparity errors, K-nearest neighbours (KNN) is used. It identifies and removes outliers based on average distances to local neighbours, enhancing point cloud accuracy [23]. Due to stereo vision's reduced accuracy at long range, the region of interest (ROI) is limited to closer areas where depth estimates are reliable. This minimises distant errors and reduces computation [19].

**Digital Elevation Model.** 3D points are transformed from the camera frame $(X_c, Y_c, Z_c)$ to the ground frame $(X_g, Y_g, Z_g)$, using extrinsic parameters that account for camera position and orientation on the rover [24]. The point cloud is then discretised into a 2D grid, a process known as voxelisation. Each cell stores the aggregated elevation; here, mean values are used, simplifying terrain data and enabling efficient processing [1]. A 2.5D digital elevation model (DEM) is generated from a single disparity map (see Fig. 6). Further processing, merge multiple DEMs into a regional terrain model and analyse terrain features based on rover capabilities [2].

**Binary Navigation Map.** The DEM is further binarised into a binary navigation map to indicate traversable areas, representing a local terrain map collected at each navigation stop [2]. This map classifies the terrain by considering hazard thresholds, such as rock height, to avoid damaging the rover's undercarriage or solar panels [1]. It supports safe path planning and slope filtering.

## 5. EXPERIMENTAL RESULTS

The proposed method has been evaluated in three datasets: the Mars-analogue dataset from Devon Island [25], the Spring benchmark [22] containing Martian-like features, and the Middlebury benchmark [5] to evaluate the method's general robustness and adaptability beyond planetary datasets. Due to the challenge of obtaining ground truth disparity maps in planetary applications, we first evaluate our method on the stereo benchmarks. Second, we experiment in Martian scenarios with varying illumination and rocky fields, to evaluate resulting terrain models and navigation maps. In all experiments, we set the constant penalties in SGM to $P_1 = 4$ and $P_2 = 38$, and test on 4000 superpixels.

**Middlebury Benchmark Evaluation.** We first tested our method on the Middlebury stereo benchmark. This widely used dataset provides a controlled environment for comparing disparity estimation accuracy. Results are evaluated by calculating the percentage of erroneous pixels, defined as disparities deviating more than 2 pixels from ground truth in both non-occluded (nonocc) and full regions (all) [5]. As shown in Table 1, our method outperforms the baseline SGM across both test scenes.

Table 1. Percentage of erroneous disparity values of proposed algorithm and comparisons.

| Algorithm | PlaytableP | | Vintage | |
|---|---|---|---|---|
| | nonocc | all | nonocc | all |
| Proposed | 12.64 | 17.64 | 22.22 | 19.42 |
| SGM | 14.6 | 18.0 | 39.3 | 43.2 |

**Spring Benchmark Evaluation.** To evaluate the performance on Martian characteristics such as less colourful, repetitive patterns, and low texture, we further test in the Spring dataset. This dataset is generated from an open-source animation with various outdoor scenarios [22]. The image resolution is 1920x1080 pixels. In this experiment, we test at half resolution and use selected training image pairs in four scenes with ground truth disparity maps. The evaluation metric is Root Mean Square (RMS) error [5] out of all pixels. Table 2 presents a quantitative result of SGM and the proposed method. It shows that the average disparity error of the proposed method is less than 2.5 pixels and can achieve an average of 0.5 pixels better than the baseline SGM among the 4 datasets. The qualitative results are shown in Fig. 4.

Table 2. RMS error of SGM and the proposed method (unit: pixels)

| Methods | | SGM [6] | Proposed |
|---|---|---|---|
| scene-02 | all | 6.89 | 3.43 |
| scene-04 | all | 0.84 | 0.54 |
| scene-22 | all | 2.35 | 1.19 |
| scene-36 | all | 2.59 | 1.69 |

**Martian Scenario.** The Devon Island rover navigation dataset [25] is collected at a Mars analogue site on Devon Island, Canada. The image resolution is 1280x960 pixels. We correct the stereo frame's point cloud by compensating for rover tilt using pitch and roll from the inclinometer. A rotation matrix is derived to align the gravity vector, followed by applying the given camera-to-inclinometer transform from the dataset.

The disparity maps and corresponding terrain models are shown in Fig. 5. The proposed method produces smooth disparity estimates and clearly distinguishes soil and rock regions, with a terrain grid resolution of 0.05 meters per pixel. Occluded areas behind rocks appear as holes in the terrain maps, particularly around large rocks in the second scene of Fig. 5. As discussed in Section 4, depth estimation degrades with distance, so an 8-meter ROI is applied, causing distant objects like background rocks to be excluded from the terrain maps. Fig. 6 shows the 2.5D voxel-based terrain model of the second scene (cropped for clarity) alongside its binary navigation map. The map applies a 0.2-meter elevation threshold, where white indicates traversable regions and black denotes obstacles.

## 6. CONCLUSIONS

The proposed approach of combining semi-global matching with superpixel-based refinement demonstrates an enhancement of the accuracy of disparity maps, successfully dealing with typically challenging low-texture regions and sharp depth discontinuities in Martian scenarios.

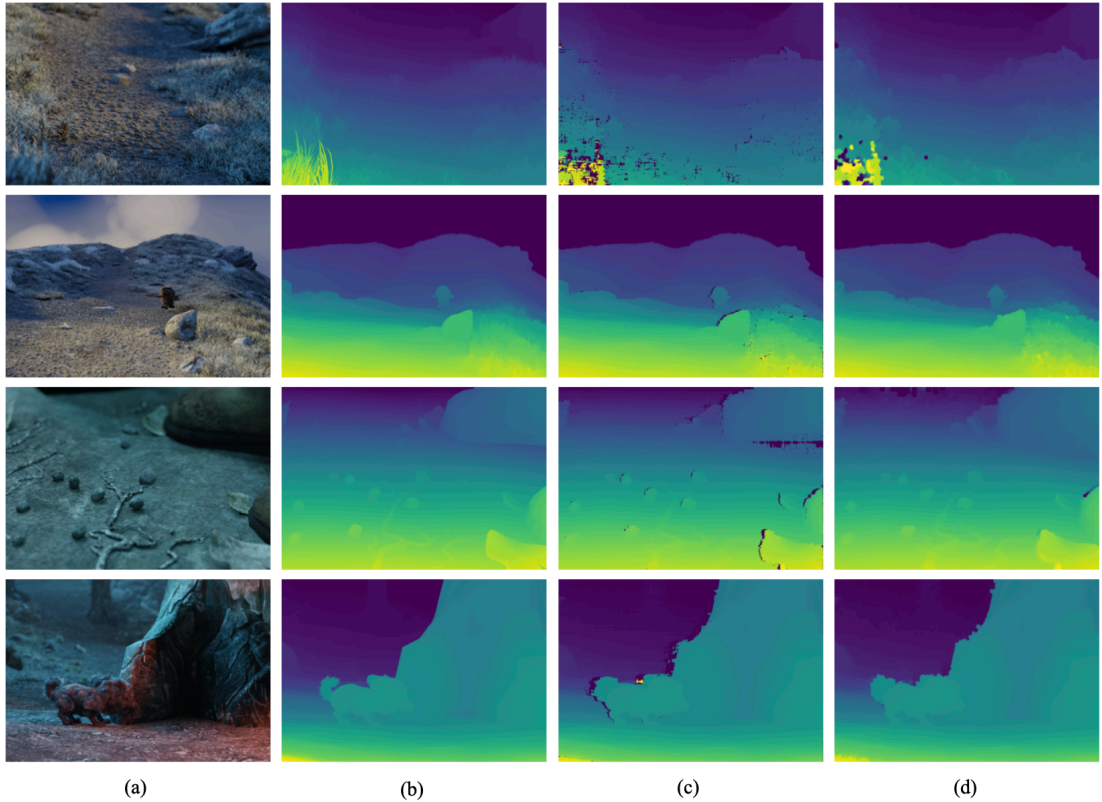

Fig. 4. Qualitative results of SGM and the proposed method. (a) reference image, (b) ground truth, (c) SGM disparity map, (d) proposed method. From top to bottom: scene-02, scene-04, scene-22, scene-36 in the Spring dataset.

This method effectively captures fine details and maintains the terrain features, leading to sharper and more precise depth estimations. This improvement allows the rover to better detect and avoid obstacles, assess terrain traversability, and plan safer paths.

**ACKNOWLEDGEMENT**

We thank Piotr Weclewski and Maria Carrillo Barrenechea for their expertise discussions on the topic.

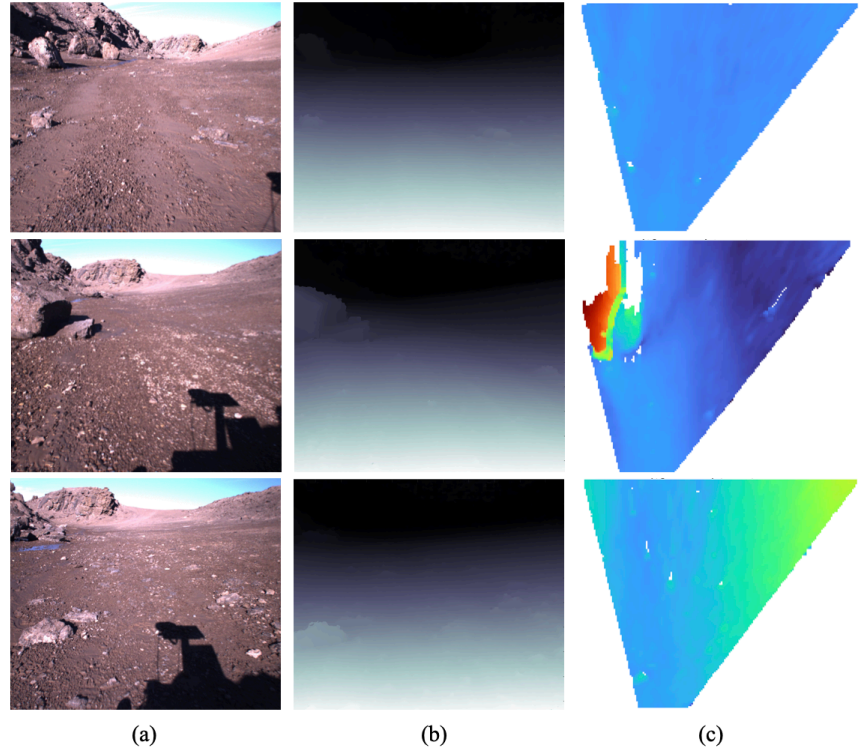

Fig. 5. Test results of three scenes in the Devon dataset. (a) Original scene, (b) refined disparity map, (c) 2D terrain map, using colour-coding to represent elevation, ranging from 0 m (dark blue) to 1.2 m (red). Occlusions behind rocks result in empty regions, visible as holes in the map. Objects in the distance are not visible because of the ROI limitation.

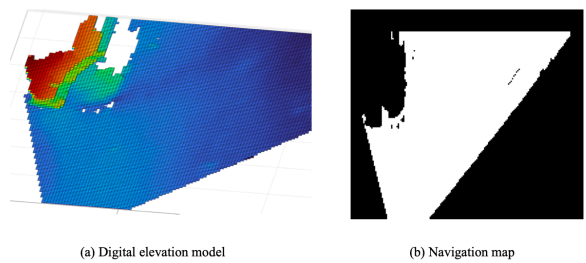

(a) Digital elevation model  (b) Navigation map

Fig. 6. Sample output of the second scene in Fig. 5. (a) 2.5D digital elevation model (DEM), (b) 2D binary navigation map.